\input{aipcheck}

\documentclass[
    ,final            
  ]
  {aipproc}

\layoutstyle{6x9}

\begin{document}

\title{Metal-rich T-dwarfs in the Hyades cluster}


\author{J. Bouvier}{
  address={Laboratoire d'Astrophysique, Observatoire de Grenoble,
  Universit\'e J. Fourier, CNRS, France }
}
\author{T. Kendall}{
  address={Centre for Astrophysics Research, Univ. Hertfordshire,
   College Lane, Hatfield AL10 9AB, UK}
}
\author{G. Meeus}{
  address={Astrophysikalisches Institut Potsdam, An der Sternwarte 16, D-14482
   Potsdam, Germany}
}

\begin{abstract}
We present the results of a search for brown dwarfs (BDs) and very low
mass (VLM) stars in the 625 Myr-old, metal-rich ([Fe/H]=0.14) Hyades
cluster. We performed a deep (I=23, z=22.5) photometric survey over 16
deg$^2$ around the cluster center. We report the discovery of the
first 2 BDs in the Hyades cluster, with a spectral type T1 and T2,
respectively. Their optical and near-IR photometry, as well as 
their proper motion, are consistent with them being cluster members.
According to models, their mass is about 50 Jupiter masses at an age
of 625 Myr. We also report the discovery of 3 new very low mass
stellar members and confirm the membership of 15 others.
\end{abstract}

\maketitle


\section{The Hyades cluster }

The Hyades (Melotte 25, $\alpha_{2000}$=04$^h$26$^m$54$^s$,
$\delta_{2000}$=+15$^o$52$'$; $l$=180.05$^o$, $b$=-22.40$^o$) is one
of the richest open clusters and the closest to the Sun. Perryman et
al. (2008) derived its main structural and kinematical properties
based on Hipparcos measurements~: a distance of 46.3$\pm$0.27~pc, an
age of 625$\pm$50~Myr, a metallicity [Fe/H] of 0.14$\pm$0.05, a
present-day total mass of about 400~M$_\odot$, a tidal radius of
10.3~pc, a core radius of 2.5-3.0~pc and negligible extinction on the
line of sight. The large proper motion of the cluster
($\mu\simeq$100~mas yr$^{-1}$) can easily be measured from imaging
surveys over a timeframe of only a few years, which helps in assessing
cluster's membership.

\section{The CFHT survey}

Wide-field optical images were obtained in the I and z bands with the
CFHT 12K camera, a mosaic of 12 CCD arrays with a pixel size of
0.21$^"$ which provides a FOV of 42$'$$\times$28$'$. The survey consists
of 53 mosaic fields covering a total of 16 square degrees. It extends
symmetrically around the cluster's center, along a 4 deg-wide stripe
of constant galactic latitude, and up to 3 degrees away from the
cluster center in galactic longitude.  The survey is at least 90\%
complete down to I$\sim$23.0 and z$\sim$22.5, a limit which varies
only slightly with seeing conditions (0.6-0.8 arcsec).

\pagebreak

\section{Candidate member selection}

\begin{figure}[t]
\setlength{\unitlength}{1cm}
\graphicspath{{figures/}}
\centering
\begin{tabular}[c]{ll}
\includegraphics[width=0.6\linewidth]{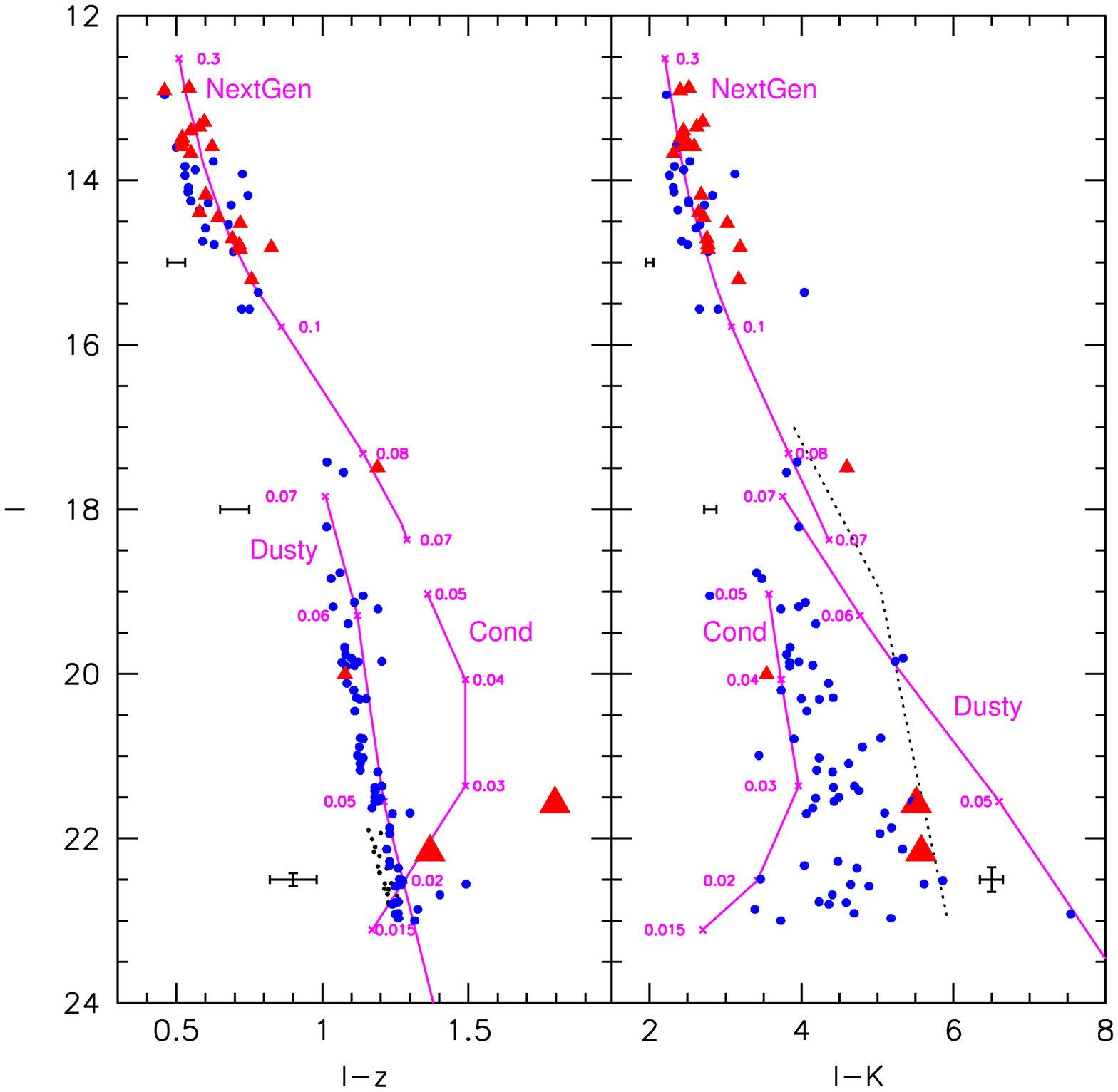} & \includegraphics[width=0.4\linewidth]{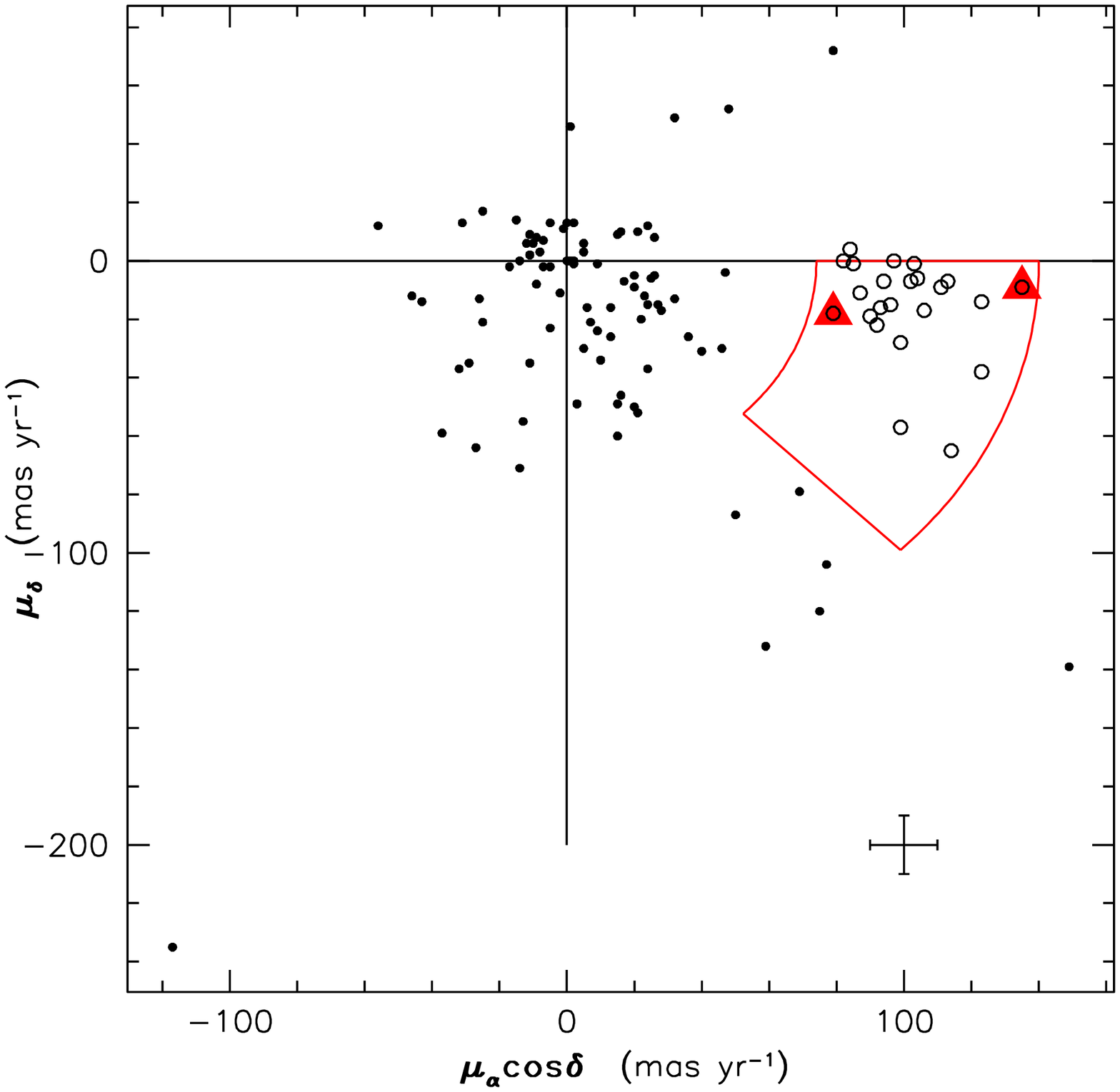}\\
\end{tabular}
\caption{{\bf Left~:} (I, I-z) and (I, I-K) CMDs of optically selected candidated
   followed up with CFHT IR in the K-band. Small dots~: 17
   optically selected candidates without follow up IR photometry. Large
   dots : optically selected candidates whose proper motion is
   inconsistent with Hyades membership (cf. right panel). Triangles :
   candidates whose proper motion is consistent with Hyades
   membership. The stellar/substellar boundary occurs at
   I$\simeq$17.8~mag. The 2 most promising substellar cluster candidates are
   shown by large triangles. NextGen (0.07-0.3~M$_\odot$), Dusty
   (0.04-0.07~M$_\odot$), and Cond (0.015-0.05~M$_\odot$) 600~Myr isochrones
   are shown and labelled with mass (Baraffe et al. 1998; Chabrier et
   al. 2000). In the (I, I-K) CMD, the dotted line indicates the locus
   of M8-T5 field dwarfs (from Dahn et al. 2002). The rms photometric
   error is shown as bars. {\bf Right~:} Proper motion vector diagram
   for 107 optically selected candidates followed up in the K-band
   (see text). The expected proper motion for Hyades members is shown
   by the (red) box (Bryja et al. 2004). Within these boundaries, 23
   optically selected candidates (empty circles) are found to share
   the proper motion of the cluster, including 2 BDs (large
   triangles). Typical rms errors on the ppm measurements are shown by
   a cross.}
\label{cmd}
\end{figure}

PSF photometry was performed on the I and z-band images with a
modified version of SExtractor (Bertin \& Arnouts 1996) from a PSF
model computed with the PSFEx software.  The
(I, I-z) color magnitude diagram (CMD) is shown in Fig~\ref{cmd}. A
total of 125 possible Hyades members were selected in this CMD from
their location relative to model isochrones. Follow up K-band imaging
was obtained for 108 of the 125 optically selected candidate members
using the 1k$\times$1k CFHT IR camera.  The (I,
I-K) CMD for the 108 candidates followed up in the K-band is shown in
Fig~\ref{cmd}. In addition, the proper motion of optically selected
Hyades candidates was computed from pairs of optical (I, z) and
infrared (K) images obtained 2 or 3 years apart. The proper motion
vector diagram of 107 optically selected Hyades candidate members is
shown in Figure~\ref{cmd}.

\begin{table}[t]
\caption{The lowest mass Hyades members : photometry and proper motion.}             
\label{good}      
\centering                          
\begin{tabular}{l l l l l l l l l l l}        
\hline\hline                 
CFHT-Hy-\# & RA(2000) & Dec(2000) & I & I-z & I-K & $\mu_{\alpha\cos\delta}$& $\mu_\delta$ &Mass\\    
&&&&&& \multicolumn{2}{c}{($mas.yr^{-1}$)}&(M$_\odot$)  \\
\hline                        
CFHT-Hy-19 &  4 17 24.8  &  16 34 36  &  17.49  &  1.18  &  4.59  &99  &  -28  & 0.08\\
CFHT-Hy-20 &  4 30 38.7  &  13 09 57  &  21.58  &  1.79  &  5.50  &135  &  -9 & 0.05  \\
CFHT-Hy-21 &  4 29 22.7  &  15 35 29  &  22.16  &  1.36  &  5.57  &79  &  -18 & 0.05  \\
\hline                                   
\end{tabular}
\end{table}

\begin{figure}[ht!]
\setlength{\unitlength}{1cm}
\graphicspath{{figures/}}
\centering
\begin{tabular}[c]{ll}
\includegraphics[width=0.6\linewidth]{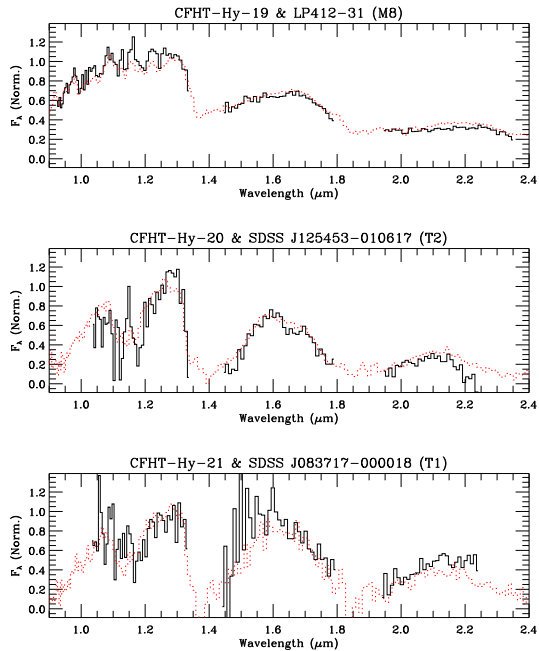} \\
\end{tabular}
\caption{Near-infrared Amici low resolution spectra of CFHT-Hy-19,
20 and 21 (solid lines from top to bottom). In each panel we also show
the closest matching field dwarf spectrum (dotted line) from the low resolution
Amici spectral library (Testi et al. 2001; Testi 2004). }
\label{nics}
\end{figure}

\section{Brown dwarfs in the Hyades cluster} 

Based on photometry and astrometry, we eventually identified 20
candidates which consistently qualify as probable Hyades members on
the basis of their optical photometry, (I-K) color and proper
motion. Of these, 15 were already listed as possible or probable
Hyades members in Prosser \& Stauffer's Open Cluster Database. The
remaining 5 probable members we report here are new. They include 3
low mass stars ($\sim$0.14~M$_\odot$) and 2 objects well within the
substellar regime ($\sim$0.050~M$_\odot$).

Our survey thus identifies the first 2 Hyades BD candidates
(CFHT-Hy-20, 21) as well as a previously detected very low mass star
(CFHT-Hy-19) close to the stellar-substellar boundary, that had
originally been considered as a non-member by Gizis et al. (1999).
The properties of these lowest mass members are listed in Table~1. The
2 BDs are well within the substellar domain with an estimated mass of
about 50 Jupiter masses while the lowest mass star has an estimated
mass around 0.08~M$_\odot$. Low resolution infrared spectra were
obtained for these 3 objects using TNG/NICS and are shown in
Figure~\ref{nics}. Fitting the observed spectra with those of template
field dwarfs observed with the same instrument, we derive a spectral
type of M8, T2 and T1 for CFHT-Hy-19, 20 and 21, respectively.

The 2 T-dwarfs we report here are strong candidate Hyades members
based on their consistent photometry and proper motion. Nevertheless,
we proceed in estimating the probability that they could be unrelated
field T dwarfs projected onto the Hyades cluster.  From the
combination of the 2MASS and SDSS DR1 surveys, Metchev et al. (2008)
derived an upper limit of 0.9$\times$10$^{-3}$ pc$^{-3}$ on the space
density of T0-T2.5 dwarfs in the solar neighborhood. Combining the
area of our survey with the range of distances for possible field
contaminants, the corresponding volume is 65~pc$^3$. We thus expect
$\leq$0.06 early field T dwarf to contaminate our survey. This
estimate further strengthens the likelihood that the 2 candidates we
report here are indeed the first BDs and the lowest mass members of
the Hyades cluster known to date.

\section{Conclusion} 

Our survey is complete in the mass range from less than 50 Jupiter
masses up to 0.20~$M_\odot$. In this mass range, we identified 18 very
low mass stars, down to the stellar-substellar limit, as well as 2
brown dwarfs with a spectral type T1 and T2. These are
the first T-dwarfs identified in the Hyades cluster at an age of
625~Myr, and also the only known instances of metal-rich ([Fe/H]=0.14)
methane dwarfs. A full account of these results is given in Bouvier et
al. (2008)\footnote{More recently, Hogan et al. (2008) reported the
discovery of 12 L-dwarfs in the cluster.}. Additional spectroscopy is
planned on Gemini during the fall of 2008 in order to investigate the
spectral characteristics of metal-rich T-dwarfs in more detail and
confront them with model predictions.



\end{document}